# Broadband radial discone antenna: Design, application and measurements

N.I. Yannopoulou, P.E. Zimourtopoulos and E.T. Sarris

The wide band of frequencies that includes all those allocated to 2G/3G applications was defined as 2G/3G band and the discone antenna with a structure of radial wires was defined as radial discone. This antenna was theoretically analysed and software simulated with the purpose of computationally design a broadband model of it. As an application, a radial discone for operation from 800 to 3000 MHz, which include the 2G/3G band, was designed and an experimental model was built and tested. Mathematically expressed measurement error bounds were computed in order to evaluate the agreement between theory and practice.

*Introduction*: Kandoian invented the well-known discone antenna in 1945 [1]. Nail gave two simple relations for discone dimensions in 1953 [2]. Rappaport designed discones using an N-type connector feed in 1987 [3]. Cooke studied a discone with a structure of radial wires in 1993 [4]. Kim et al. recently presented a double radial discone antenna for UWB applications [5]. This Letter proposes a broadband radial discone fed by an N-type connector for operation from 800 to 3000 MHz that includes all frequencies allocated to 2G/3G applications.

*Analysis, Simulation, Design, Application and Measurements*: The radial discone was theoretically analysed as a group of identical filamentary V-dipoles with unequal arms connected in parallel. The dipoles recline on equiangular vertical phi-planes around z-axis to form a disc-conical array. Fig. 1-A shows two coplanar dipoles conformed to the apex angle (a). Each dipole has a total length L equal to the sum of arm lengths r and s and the gap between its terminals g.

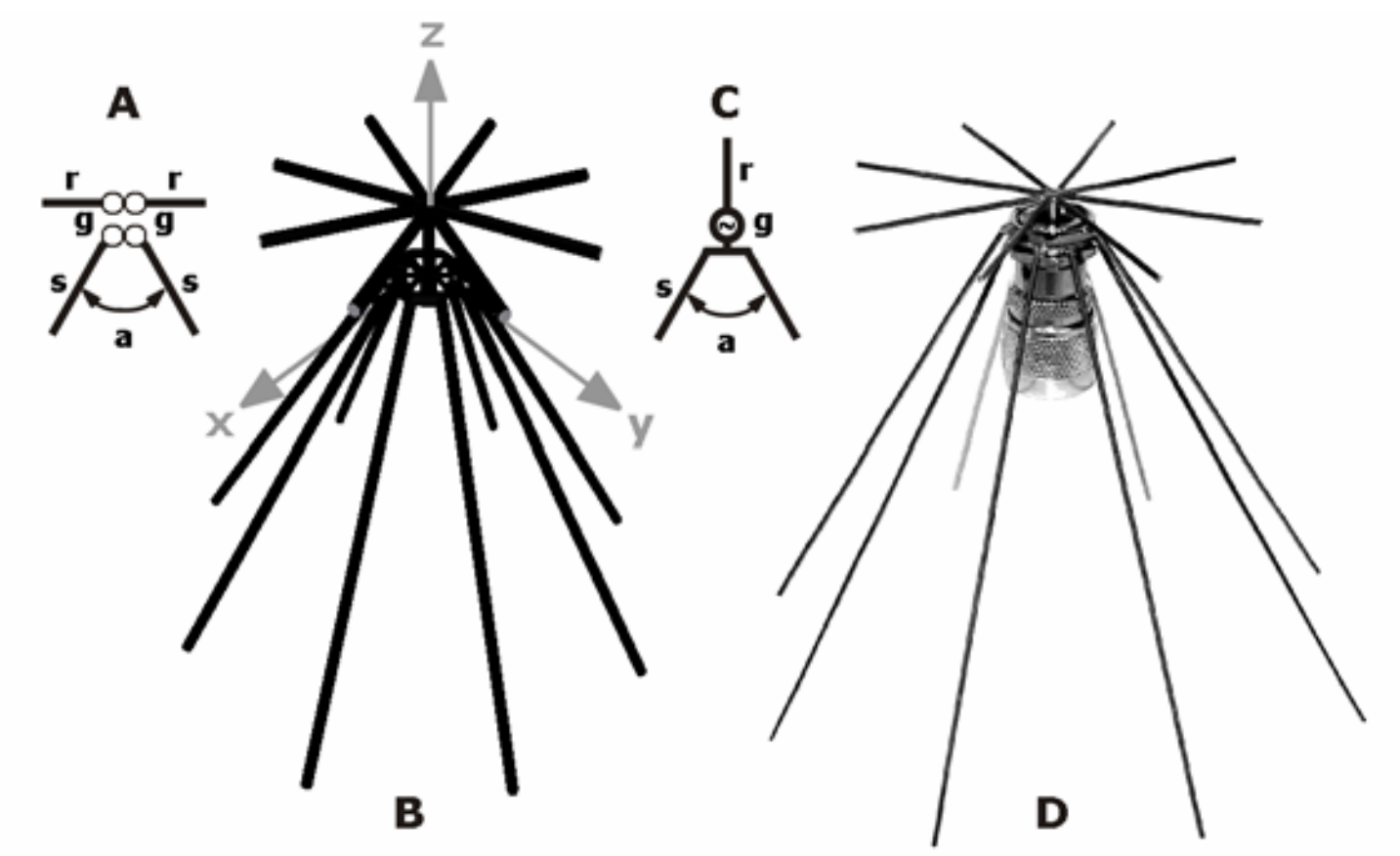

Fig. 1 Radial Discone models - Theoretical (A), Designed (B) and Built (D) - and the reference Ground Plane antenna (C)

The simulation was based on a suite of developed visual tools supported by a fully analysed, corrected and redeveloped edition of the original thin-wire computer program by Richmond [6].

A visual program was specifically developed to design a broadband radial discone with bare wires of diameter d embedded in free space when the wire conductivity, the type of feeding connector and the frequency band are given. Two arithmetic criteria were adopted for the broadband characterisation of a model: VSWR lower than 2 and normalized radiation intensity U/Umax lower than 3 dB on horizontal plane. The program uses the model of a radial discone fed by an N-type connector shown in Fig. 1-B. Starting with an appropriate combination of the relations given by [2]-[4] the program computes by iteration in terms of wavelength $\lambda$, the characteristics r, s, g, (a) and d of a broadband model, just when the criteria are satisfied. Fig. 1-C shows a ground plane antenna consists of the same cone radials and a vertical monopole with height r, which was designed for reference.

As a practical application of the broadband design, the 2G/3G band from 800 to 2500 MHz was selected to begin with and an experimental radial discone of copper wire fed by N-type connector was built, as shown in Fig. 1-D. In order to demonstrate the particular behaviour of the experimental model, the 2G/3G band was divided in six sub-bands I: 806-960 MHz, II: 1429-1513 MHz, III: 1710-1900 MHz, IV: 1910-2025 MHz, V: 2110-2170 MHz and VI: 2400-2499 MHz.

Our measurement system consists of an EM anechoic chamber, a network analyser, a number of support instruments, a set of standard loads of factory accuracy and a constructed antenna rotation mechanism with a built hardware control unit of its step motor. The combined characteristics of system parts specify a measurement band from 600 to 1300 MHz, which overlaps with the 2G/3G band. Developed control software synchronises the system and collects data using the IEEE-488 protocol. A developed general mathematical method expresses the measurement error bounds. Another set of developed software applications processes the collected data and computes the error bounds.

*Results*: The consideration of radial discone as an array of at least 8 V-dipoles produces a theta-polarised vector radiation pattern with magnitude a surface almost by revolution around z-axis. So the radial discone has indeed on horizontal plane xOy the basic properties of a vertically polarised almost omni-directional antenna, a fact that encouraged design of a broadband model by simulation.

The application of the broadband criteria to 2G/3G band resulted to the design of a radial discone with r = 44 mm, g = 6 mm, s = 125 mm, d = 1.5 mm and a = 60° that operates from 800 to 3000 MHz, which exceeds that of 2G/3G band. The accordingly built experimental radial discone of Fig. 1-D should be implied with a constructional tolerance of ±0.5 mm and ±0.5°.

The broadband model has a directivity from about -0.5 to 2.9 dBd with slope angle between -65° and +58°, but the directivity gain on horizontal plane stays very close to the desirable value of 0 dBi, since it changes from -1 to +1.7 dBi only. Fig. 2 shows that the predicted horizontal normalised radiation intensity remains below 3 dB indeed, while it stays above 0 dB relative to the reference antenna in all 2G/3G sub-bands indicated by the vertical grey strips, when both are fed by the same 50-Ohm source.

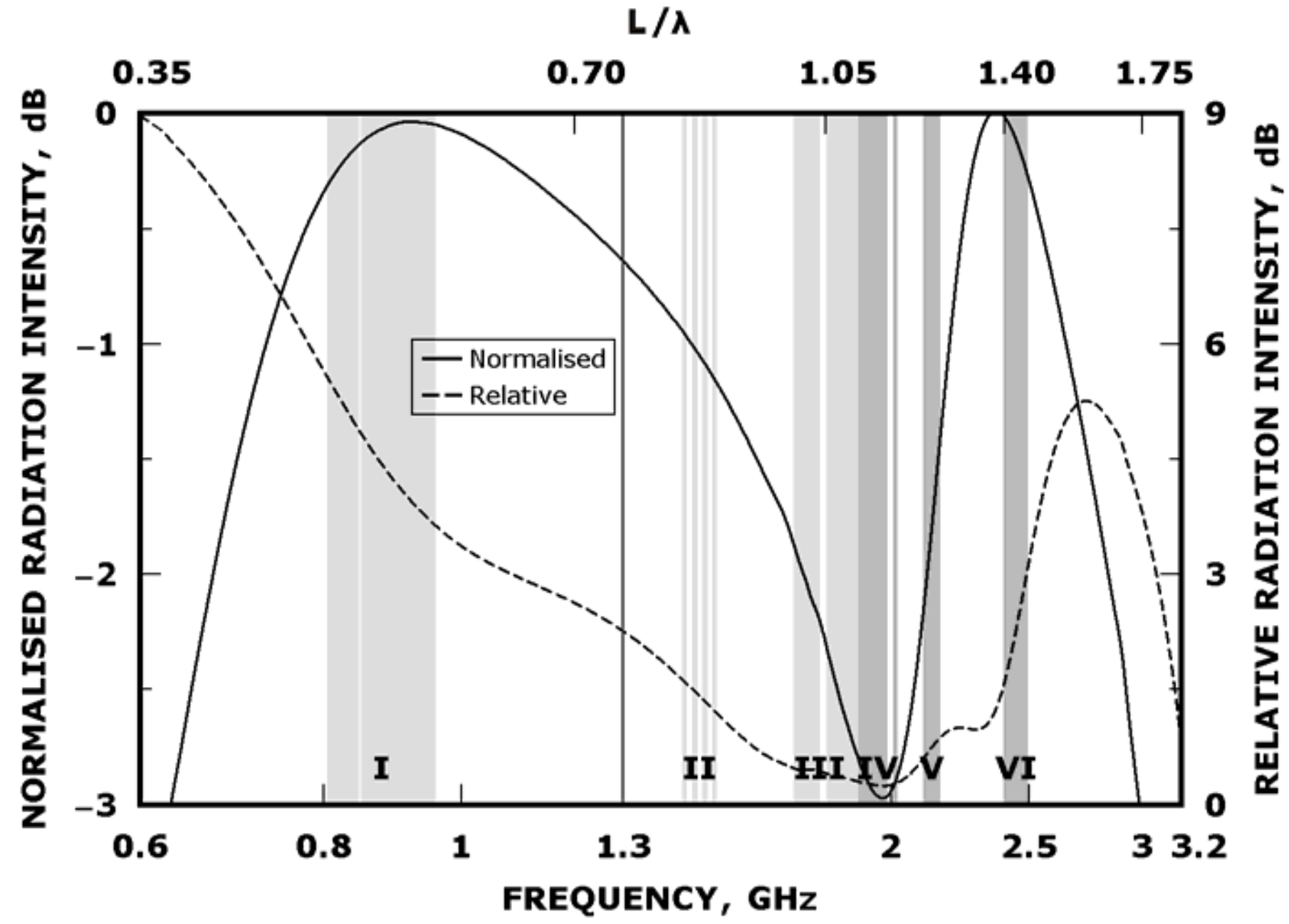

Fig. 2 Predicted normalised and relative radiation intensity on horizontal plane against frequency or ratio of total length to wavelength

Fig. 3 shows the predicted normalized radiation patterns in dB at the centre of each sub-band, which confirms the horizontal omni-directional radiation properties of the broadband model.

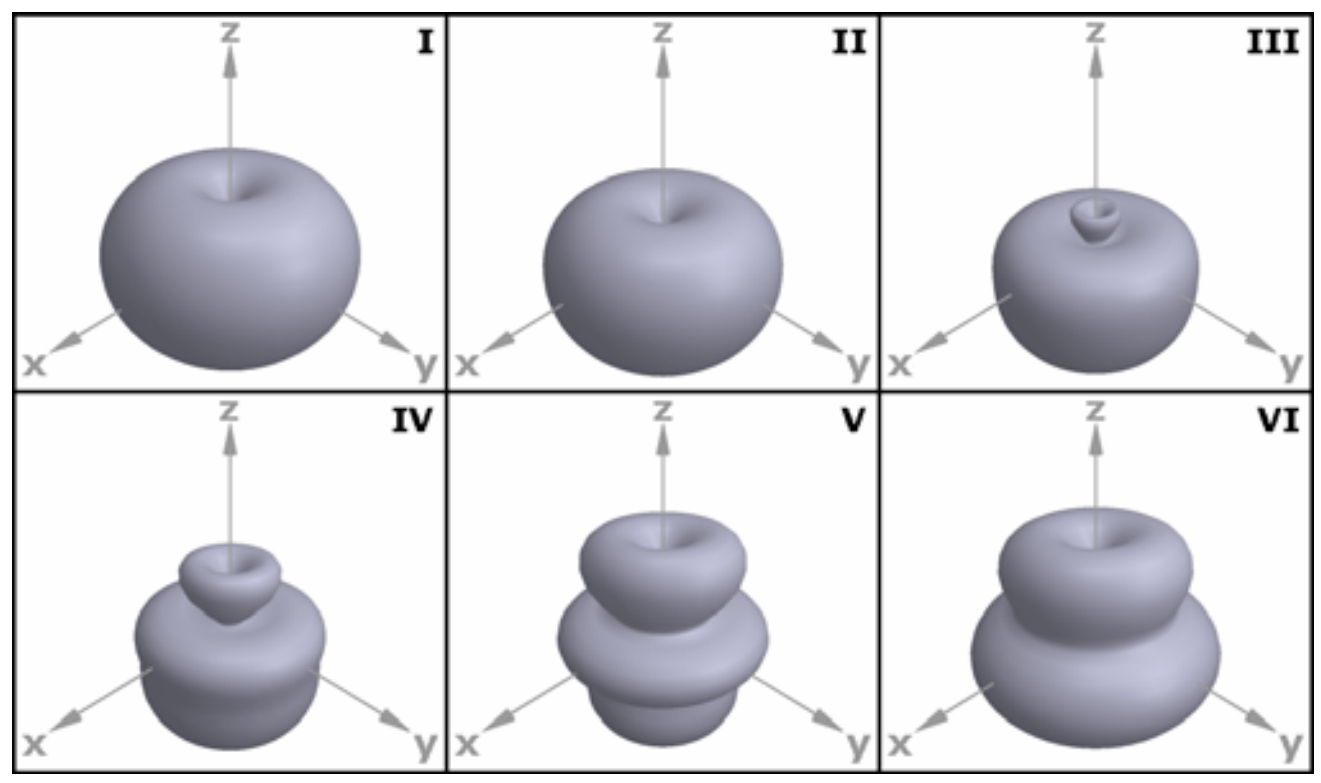

Fig. 3 Predicted normalised radiation intensity patterns

at the centre of each 2G-3G sub-band

At 950 MHz, the centre frequency of the measurement band, the predicted and measured radiation intensity on the three main cuts of the radiation pattern are in good agreement, as shown in Fig. 4.

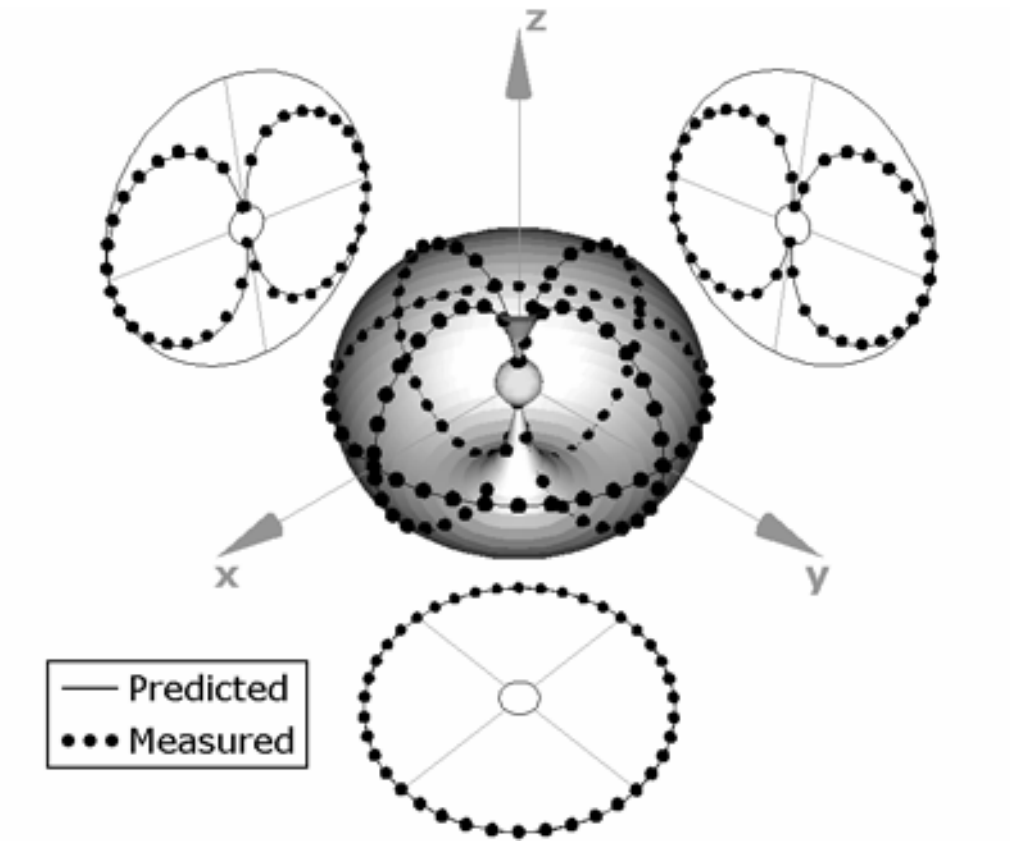

Fig. 4 Normalised radiation intensity pattern at the centre of measurements band

This is made clearer by the measurement error bounds on a vertical plane, as shown in Fig. 5.

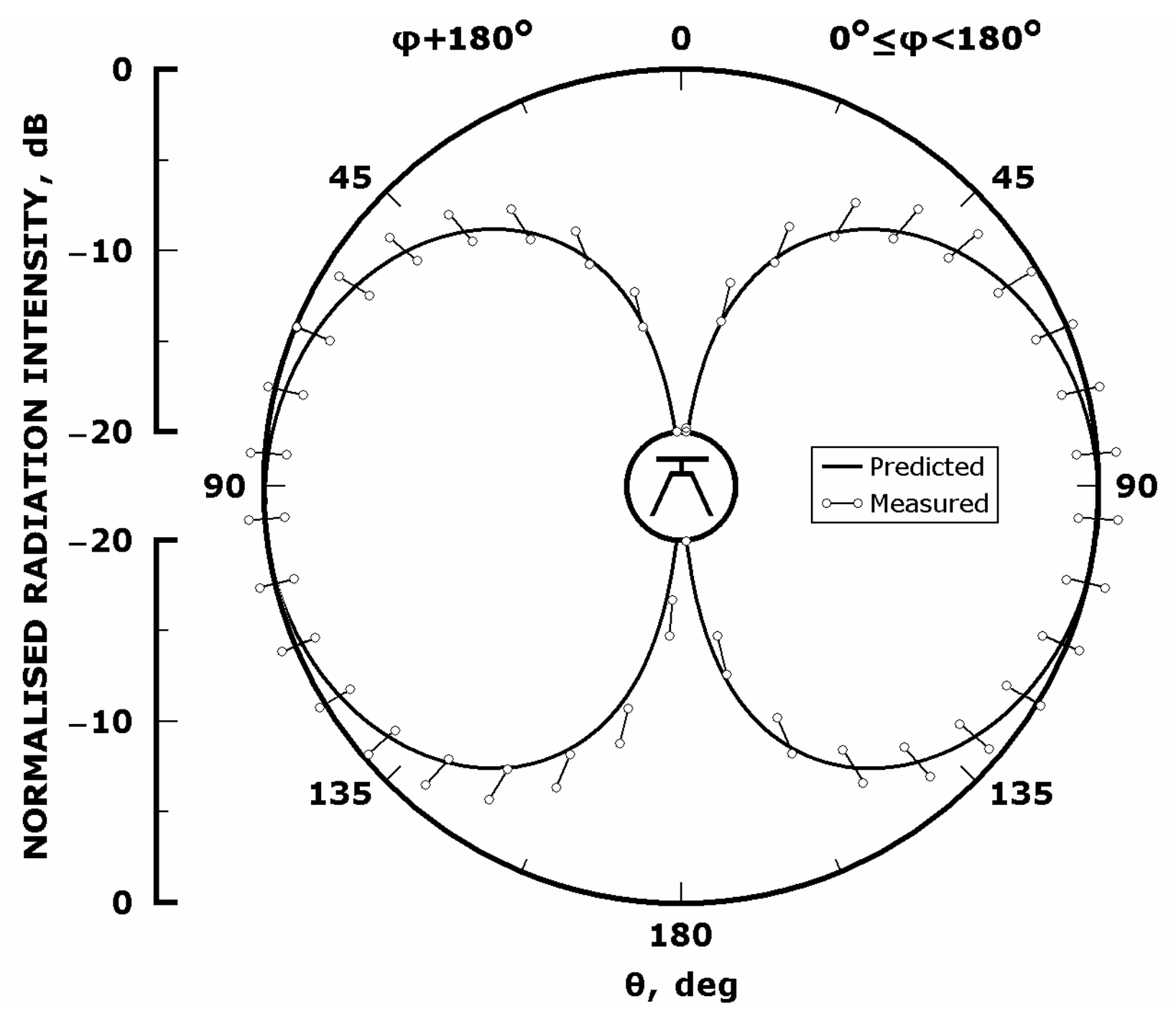

Fig. 5 Normalised radiation intensity pattern on a vertical plane

at the centre of measurements band

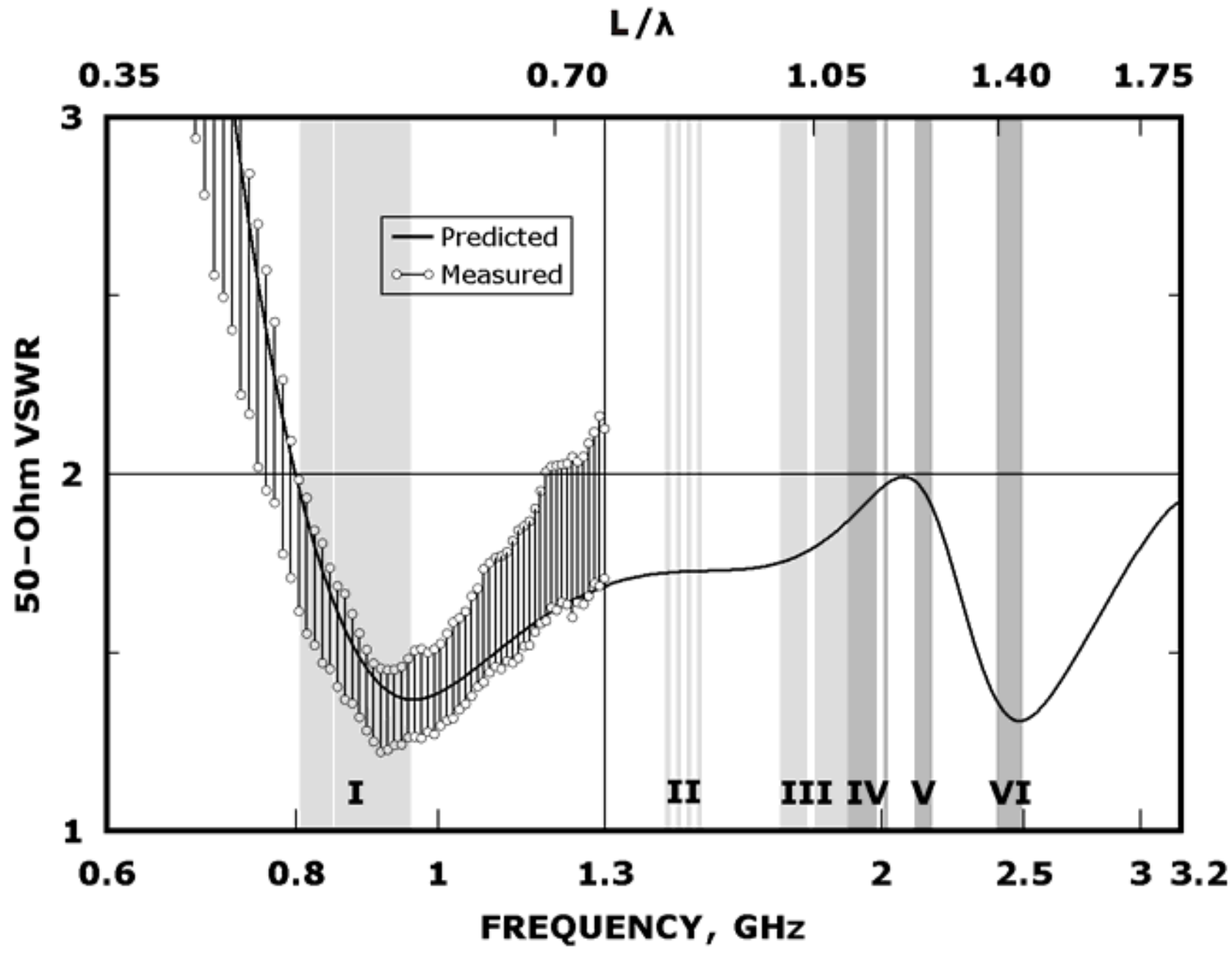

Fig. 6 Standing wave ratio against frequency or ratio of total length to wavelength

Fig. 6 shows that the 50-Ohm VSWR predicted results for the broadband discone are below 2 indeed and almost covered by the error bounds in the measurement band.

Therefore, prediction and experimentation in the measurement band proposes a successfully designed broadband radial discone antenna for operation from 800 to 3000 MHz.

N.I. Yannopoulou, P.E. Zimourtopoulos and E.T. Sarris (Section of Telecommunication and Space Science, Department of Electrical and Computer Engineering, Democritus University of Thrace, V.Sofias 12, Xanthi, 671 00, Greece)